\documentclass[fleqn]{llncs}

\usepackage{verbatim}

\usepackage{amsmath}
\usepackage{amssymb}
\usepackage{stmaryrd} 
\usepackage{wasysym}
\usepackage{mathtools} 
\usepackage{bm}
\usepackage{listings}
\usepackage{xcolor}
\usepackage{pifont}
\usepackage{multicol}
\usepackage{marvosym}
\usepackage[]{algorithm2e}
\usepackage{pgfplots}
\pgfplotsset{compat=1.12}
\definecolor{vgreen}{RGB}{104,180,104}
\definecolor{vblue}{RGB}{49,49,255}
\definecolor{vorange}{RGB}{255,143,102}

\lstdefinestyle{verilog-style}
{
	language=Verilog,
	basicstyle=\small\ttfamily,
	keywordstyle=\color{vblue},
	identifierstyle=\color{black},
	commentstyle=\color{vgreen},
	numbers=left,
	numberstyle=\tiny\color{black},
	numbersep=10pt,
	tabsize=8,
	moredelim=*[s][\colorIndex]{[}{]},
	literate=*{:}{:}1
}
\makeatletter
\newcommand*\@lbracket{[}
\newcommand*\@rbracket{]}
\newcommand*\@colon{:}
\newcommand*\colorIndex{%
	\edef\@temp{\the\lst@token}%
	\ifx\@temp\@lbracket \color{black}%
	\else\ifx\@temp\@rbracket \color{black}%
	\else\ifx\@temp\@colon \color{black}%
	\else \color{vorange}%
	\fi\fi\fi
}

\usepackage{cite}
\usepackage{tikz}
\usetikzlibrary{arrows,automata,positioning}

\usepackage{ltltikz}

\usepackage{subfig}
\usepackage{wrapfig}

\usepackage[firstpage]{draftwatermark}\SetWatermarkText{\hspace*{5.2in}\raisebox{5in}{\includegraphics[scale=0.1]{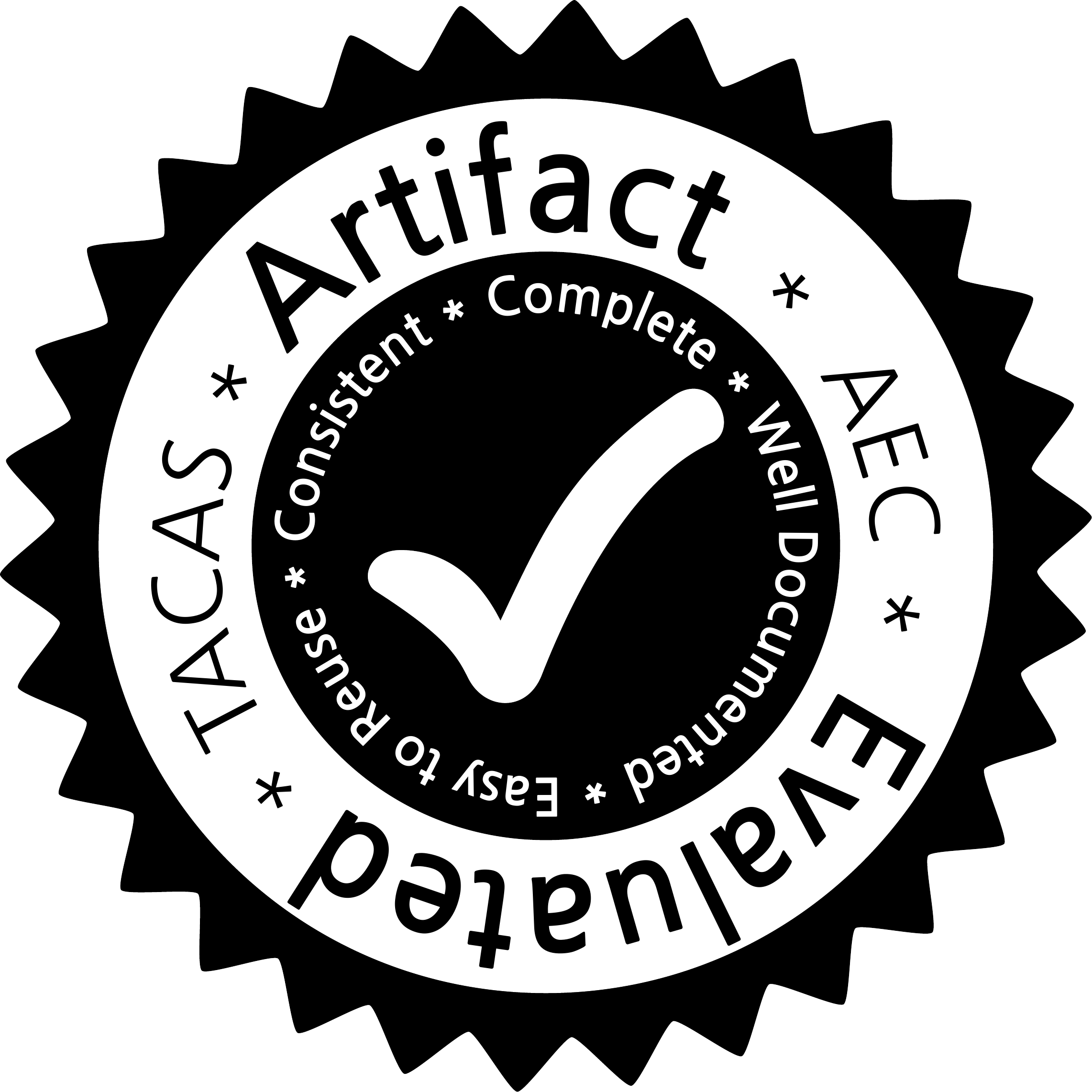}}}\SetWatermarkAngle{0}

\newcommand{\set}[1]{\{#1\}}

\newcommand{\ldot}{\mathpunct{.}}

\newcommand{\ltl}{\text{LTL}}
\newcommand{\hyperltl}{\text{HyperLTL}}

\newcommand{\lang}{\mathcal{L}}
\newcommand{\ap}{\text{AP}}
\renewcommand{\models}{\vDash}
\newcommand{\nmodels}{\nvDash}

\newcommand{\U}{\Until}
\newcommand{\X}{\Next}

\newcommand{\W}{\WUntil}

\newcommand{\pathvars}{\mathcal{V}}
\newcommand{\pathassignfin}{\Pi_\mathit{fin}}

\newcommand{\monitor}{\mathcal{M}}
\newcommand{\tool}{\text{RVHyper}}

\newcommand{\eahyper}{\text{EAHyper}}

\newcommand{\ninfluences}{\not\leadsto}

\title{$\tool$: A Runtime Verification Tool for Temporal Hyperproperties\thanks{This work was partially supported by the German Research Foundation (DFG) as part of the Collaborative Research Center ``Methods and Tools for Understanding and Controlling Privacy'' (SFB 1223) and by the European Research Council (ERC) Grant OSARES (No. 683300).}}
\author{Bernd Finkbeiner \and Christopher Hahn \and Marvin Stenger \and Leander Tentrup}
\institute{Reactive Systems Group\\Saarland University\\\email{lastname@react.uni-saarland.de}}

\begin{document}

\maketitle

\begin{abstract}
	

  We present $\tool$, a runtime verification tool for hyperproperties.
  Hyperproperties, such as non-interference and observational determinism, relate multiple computation traces with each other.
  Specifications are given as formulas in the temporal logic $\hyperltl$, which extends linear-time temporal logic~(LTL) with trace quantifiers and trace variables.
  $\tool$ processes execution traces sequentially until a violation of the specification is detected.
  In this case, a counter example, in the form of a set of traces, is returned.
  As an example application, we show how $\tool$ can be used to detect spurious dependencies in hardware designs.
  
\end{abstract}

\section{Introduction}
\label{intro}
\sloppy

\emph{Hyperproperties}~\cite{journals/jcs/ClarksonS10} generalize trace properties in that they not only check the correctness of \emph{individual} computation traces in isolation, but relate \emph{multiple} computation traces to each other.
$\hyperltl$~\cite{conf/post/ClarksonFKMRS14} is a logic for expressing temporal hyperproperties, by extending linear-time temporal logic with \emph{explicit} trace quantification.
$\hyperltl$ has been used to specify a variety of information-flow and security properties.
Examples include classical properties like non-interference and observational determinism, as well as quantitative information-flow properties, symmetries in hardware designs, and formally verified error correcting codes~\cite{conf/cav/FinkbeinerRS15}. While model checking and satisfiability checking tools for HyperLTL already exist~\cite{conf/cav/FinkbeinerRS15,conf/concur/FinkbeinerH16}, the \emph{runtime verification} of HyperLTL specifications has so far, despite recent theoretical progress~\cite{conf/csfw/AgrawalB16,DBLP:conf/tacas/BrettSB17,conf/rv/FinkbeinerHST17}, not been supported by practical tool implementations.

Monitoring hyperproperties is difficult: in principle, the monitor not only needs to process every observed trace, but must also \emph{store} every trace observed so far, so that future traces can be compared with the traces seen so far. On the other hand, a runtime verification tool for hyperproperties is certainly useful, in particular if the implementation of a security critical system is not available. Even without access to the source code, monitoring the observable execution traces still detects insecure information flow.

In this paper, we present $\tool$, a runtime verification tool for monitoring temporal hyperproperties.
$\tool$ tackles this challenging problem by implementing two major optimizations: (1) a \emph{trace analysis}, which detects all redundant traces that can be omitted during the monitoring process and (2) a \emph{specification analysis} to detect exploitable properties of a hyperproperty, such as \emph{symmetry}.

We have applied $\tool$ in classical information-flow security, such as checking for violations of observational determinism. $\hyperltl$ is, however, not limited to security policies. As an example of such an application beyond security, we show how $\tool$ can be used to detect spurious dependencies in hardware designs.


\section{$\tool$}
In this section we give an overview  on the monitoring approach, including the input and output of the monitoring algorithm and the two major optimization techniques implemented in $\tool$.

\subsubsection{Specification.}
The input to $\tool$ is a $\hyperltl$ specification. $\hyperltl$~\cite{conf/post/ClarksonFKMRS14} is a temporal logic for specifying hyperproperties.
The logic extends $\ltl$ with quantification over trace variables $\pi$ and a method to link atomic propositions to specific traces.
The set of trace variables is $\pathvars$.
Formulas in $\hyperltl$ are given by the grammar
\begin{align*}
\varphi &{}\Coloneqq \forall\pi\ldot\varphi \mid \exists\pi\ldot\varphi \mid \psi \enspace, \text{ and}\\
\psi &{}\Coloneqq a_\pi \mid \neg\psi \mid \psi\lor\psi \mid \X\psi \mid \psi\U\psi \enspace,
\end{align*}
where $a \in \ap$ and $\pi \in \pathvars$.
%
The finite trace semantics~\cite{DBLP:conf/tacas/BrettSB17} for $\hyperltl$ is based on the finite trace semantics of LTL.
In the following, when using $\lang(\varphi)$ we refer to the finite trace semantics of a $\hyperltl$ formula $\varphi$.
Let $t$ be a finite trace, $\epsilon$ denotes the empty trace, and $|t|$ denotes the length of a trace. Since we are in a finite trace setting, $t[i,\ldots]$ denotes the subsequence from position $i$ to position $|t|-1$.
Let $\pathassignfin : \pathvars \rightarrow \Sigma^*$ be a partial function mapping trace variables to finite traces. We define $\epsilon[0]$ as the empty set.
$\pathassignfin[i, \ldots]$ denotes the trace assignment that is equal to $\pathassignfin(\pi)[i,\ldots]$ for all $\pi$. We define a subsequence of $t$ as follows.
\[
t[i,j] = \begin{cases}
\epsilon & \text{if } i \geq |t|\\
t[i,\textit{min}(j,|t|-1)], & \text{otherwise}
\end{cases}
\]
\begin{equation*}
\begin{array}{ll}
\pathassignfin \models_T a_\pi         \qquad \qquad & \text{if } a \in \pathassignfin(\pi)[0] \\
\pathassignfin \models_T \neg \varphi              & \text{if } \pathassignfin \nmodels_T \varphi \\
\pathassignfin \models_T \varphi \lor \psi         & \text{if } \pathassignfin \models_T \varphi \text{ or } \pathassignfin \models_T \psi \\
\pathassignfin \models_T \X \varphi                & \text{if } \pathassignfin[1,\ldots] \models_T \varphi \\
\pathassignfin \models_T \varphi\U\psi             & \text{if } \exists i \geq 0 \ldot \pathassignfin[i,\ldots] \models_T \psi \land \forall 0 \leq j < i \ldot \pathassignfin[j,\ldots] \models_T \varphi \\
\pathassignfin \models_T \exists \pi \ldot \varphi & \text{if there is some } t \in T \text{ such that } \pathassignfin[\pi \mapsto t] \models_T \varphi
\end{array}
\end{equation*}
For example, above mentioned observational determinism can be formalized as the HyperLTL formula
$\forall \pi \ldot \forall \pi' \ldot (O_\pi = O_{\pi'}) \WUntil (I_\pi \not = I_{\pi'})$, where $\WUntil$ is the weak version of $\Until$.


\subsubsection{Input and Output.}
The input of $\tool$ consists of a HyperLTL formula and the observed behavior of the system under consideration.
The observed behavior is represented as a trace set $T$, where each $t \in T$ represents a previously observed execution of the system under consideration.
If $\tool$ detects that the system violates the hyperproperty, it outputs a counter example, i.e, a $k$-ary tuple of traces, where $k$ is the number of quantifiers in the HyperLTL formula.

\subsubsection{Monitoring Algorithm.}
Given a HyperLTL formula $\varphi$ and a trace set $T$, $\tool$ processes a fresh trace under consideration as depicted in Algorithm~\ref{alg:offline}. The algorithm revolves around a \emph{monitor-template} $\monitor_\varphi$, which is constructed from the HyperLTL formula $\varphi$. The basic idea of the monitor template is that it still contains every trace variables of $\varphi$, which can be initialized with explicit traces at runtime. This way, the automaton construction of the monitor template is constructed only once as a preprocessing step.

$\tool$ initializes the monitor template for each $k$-ary combination of traces in $T\cup\{t\}$. If one tuple violates the hyperproperty, $\tool$ returns that $k$-ary tuple of traces as a counter example, otherwise $\tool$ returns \emph{satisfied}.
\begin{figure}[t]
	\centering
	\begin{minipage}[t]{0.5\textwidth}
			\begin{algorithm}[H]
				\SetKwInOut{Input}{input}
				\SetKwInOut{Output}{output}
				\SetAlgoLined
				\Input{$\forall^n$ HyperLTL formula $\varphi$,\\
					set of traces $T$,\\
					fresh trace $t$							}
				\Output{satisfied or $n$-ary tuple\\witnessing violation}
				\BlankLine
				$\monitor_\varphi =$ \texttt{build\_template($\varphi$)}\;
				\BlankLine
				\For{each tuple $N \in (T\cup\{t\})^n$}{
					\eIf{$\monitor_\varphi$ accepts $N$}{
						proceed\;
					}{
						\Return $N$\;
					}
				}
				\Return satisfied\;
				\BlankLine
				\caption{A high-level sketch of the monitoring algorithm for $\forall^n$ HyperLTL formulas.}
				\label{alg:offline}
			\end{algorithm}
		\end{minipage}\begin{minipage}[t]{0.5\textwidth}
		\begin{algorithm}[H]
			\label{alg_minimization}
			\SetKwInOut{Input}{input}
			\SetKwInOut{Output}{output}
			\SetAlgoLined
			\Input{$\hyperltl$ formula $\varphi$,
				redundancy free trace set $T$,
				fresh trace $t$}
			\Output{redundancy free set of traces $T_\mathit{min} \subseteq T \cup \set{t}$}
			\BlankLine
			$\monitor_\varphi =$ \texttt{build\_template($\varphi$)}
			\BlankLine
			\ForEach{$t' \in T$}{
				\If{ $t'$ dominates $t$ }
				{
					return $T$
				}
			}
			\ForEach{$t' \in T$}{
				\If{$t$ dominates $t'$}
				{
					$T \coloneqq T \setminus \set{t'}$
				}
			}
			\Return $T \cup \set{t}$
			\BlankLine
			\caption{Trace analysis algorithm to minimize trace storage.}
			\label{alg:traceopti}
		\end{algorithm}
	\end{minipage}
	\vspace{-10pt}
\end{figure}


\subsubsection{Trace Analysis: Minimizing Trace Storage.}
\label{sec:minimizingtracestorage}
The main obstacle in monitoring hyperproperties is the potentially unbounded space consumption.
$\tool$ uses a \emph{trace analysis} to detect redundant traces, with respect to a given HyperLTL formula, i.e., traces that can be safely discarded without losing any information and without losing the ability to return a counter example.

$\tool$'s trace analysis is based on the definition of trace redundancy: we say a fresh trace $t$ is $(T, \varphi)$-redundant, if $T$ is a model of $\varphi$ if and only if $T \cup \{t\}$ is a model of $\varphi$.
The idea, depicted in Algorithm~\ref{alg:traceopti}, is to check if another trace $t'$ contains at least as much informations as $t$: we say a $t'$ dominates $t$ if $\bigwedge_{\pi \in \pathvars} \lang(\monitor_\varphi[t'/\pi]) \subseteq \lang(\monitor_\varphi[t/\pi])$.
For a fresh incoming trace, RVHyper performs this language inclusion check in both directions in order to compute the minimal trace set that must be stored to monitor the hyperproperty under consideration.

\subsubsection{Specification Analysis: Decreasing Running Time.}
$\tool$ uses a \emph{specification analysis}, which is a preprocessing step that analyzes the HyperLTL formula under consideration.
$\tool$ detects whether a formula is (1) \emph{symmetric}, i.e., we halve the number of instantiated monitors, (2) \emph{transitive}, i.e, we reduce the number of instantiated monitors to two, or (3) \emph{reflexive}, i.e., we can omit the self comparison of traces~\cite{conf/rv/FinkbeinerHST17}.

\emph{Symmetry} is especially interesting because many information flow policies satisfy this property. Consider, for example, observational determinism: 
$
\forall \pi\ldot \forall \pi'\ldot (O_\pi = O_{\pi'}) \W (I_\pi \neq I_{\pi'}).
$
$\tool$ detects symmetry by translating this formula to a formula that is unsatisfiable if there exists no pair of traces which violates the symmetry condition:
$
\exists \pi\ldot \exists \pi'\ldot \big((O_\pi = O_{\pi'}) \W (I_\pi \neq I_{\pi'})\big) \nleftrightarrow \big((O_{\pi'} = O_\pi) \W (I_{\pi'} \neq I_\pi)\big)
$.
If the resulting formula turns out to be unsatisfiable, $\tool$ omits the symmetric instantiations of the monitor automaton, which turns out to be, especially in combination with $\tool$s \emph{trace analysis}, a major optimization in practice~\cite{conf/rv/FinkbeinerHST17}.



\subsubsection{Implementation.}

$\tool$\footnote{The implementation is available at \url{https://react.uni-saarland.de/tools/rvhyper/}.} is written in C\nolinebreak[4]\hspace{-.05em}\raisebox{.4ex}{\relsize{-3}{\textbf{++}}}.
It uses \emph{spot} for building the deterministic monitor automata and the \emph{Buddy} BDD library for handling symbolic constraints.
We use the $\hyperltl$ satisfiability solver $\eahyper$~\cite{conf/cav/FinkbeinerHS17,conf/concur/FinkbeinerH16} to determine whether the input formula is reflexive, symmetric, or transitive.
Depending on those results, we omit redundant tuples in the monitoring algorithm. 



\section{Detecting Spurious Dependencies in Hardware Designs}


While $\hyperltl$ has been applied to a range of domains, including security and information flow properties, we focus in the following  on a classical verification problem, the independence of signals in hardware designs.
%
We demonstrate how $\tool$ can automatically detect such dependencies from traces generated from hardware designs.

\begin{wrapfigure}[13]{r}{0.3\textwidth}
  \centering
  \begin{tikzpicture}[auto,>=stealth',shorten >=1pt,thick,transform shape, scale=1]
  
  \node[draw,minimum size=20pt] (mux) {\textsc{mux}};
  \node[draw,fill,below=0.5 and -0.3 of mux,minimum size=20pt] (bb) {};
  \node[draw,below=0.5 and -0.3 of bb,minimum size=20pt] (imux) {\textsc{imux}};
  
  \node[above left=0.5 and 0 of mux] (i) {$i$};
  \node[above right=0.5 and 0 of mux] (iPrime) {$i'$};
  \node[left=0.5 of i] (sel) {$\mathit{sel}$};
  
  \node[below left=0.5 and -0.1 of imux] (o) {$o$};
  \node[below right=0.5 and -0.1 of imux] (oPrime) {$o'$};  
  
  \coordinate[xshift=-5pt] (muxi1) at (mux.north);
  \coordinate[xshift=5pt] (muxi2) at (mux.north);
  
  \coordinate[xshift=-5pt] (imuxo1) at (imux.south);
  \coordinate[xshift=5pt] (imuxo2) at (imux.south);
  
  \draw (i) -| (muxi1)
        (iPrime) -| (muxi2)
        (sel) |- (mux)
        (mux) -- (bb)
        (bb) -- (imux)
        (sel) |- (imux)
        (imuxo1) |- (o)
        (imuxo2) |- (oPrime)
        ;

\end{tikzpicture}
  \caption{\textsc{mux} circuit with black box}
  \label{fig:example-circuit}
\end{wrapfigure}
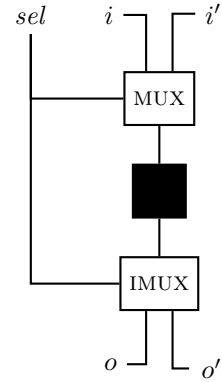

\subsubsection{Input \& Output.}
The input to $\tool$ is a set of traces where the propositions match the atomic propositions of the $\hyperltl$ formula.
For the following experiments, we generate a set of traces from the Verilog description of several example circuits by random simulation.
If a set of traces violates the specification, $\tool$ returns a counter example.

\subsubsection{Specification.}
We consider the problem of detecting whether input signals influence output signals in hardware designs.
We write $\vec i \ninfluences \vec o$ to denote that the inputs $\vec i$ do not influence the outputs $\vec o$.
Formally, we specify this property as the following $\hyperltl$ formula:
\begin{equation*}
  \forall \pi_1 \forall \pi_2 \ldot
  (\vec{o}_{\pi_1} = \vec{o}_{\pi_2}) \WUntil (\overline{\vec{i}}_{\pi_1} \neq \overline{\vec{i}}_{\pi_2}) \enspace,
\end{equation*}
where $\overline{\vec{i}}$ denotes all inputs except $\vec i$.
Intuitively, the formula asserts that for every two pairs of execution traces $(\pi_1,\pi_2)$ the value of $\vec{o}$ has to be the same until there is a difference between $\pi_1$ and $\pi_2$ in the input vector $\overline{\vec{i}}$, i.e., the inputs on which $\vec o$ may depend.

\subsubsection{Sample Hardware Designs.}
\begin{figure}[t]
	\begin{multicols}{2}
		
		\begin{lstlisting}[style={verilog-style}]
module counter(increase,
  decrease, overflow);
input increase;
input decrease;
output overflow;

reg[2:0] counter;

assign overflow = (counter
  == 3'b111 && increase
  && !decrease);


initial
begin
  counter = 0;
end
always @($global_clock)
begin
if (increase && !decrease)
  counter = counter + 1;
else if (!increase && decrease
         && counter > 0)
  counter = counter - 1;
else 
  counter = counter;
end
endmodule
		
		\end{lstlisting}
		
	\end{multicols}
	\caption{Verilog description of Example~\ref{ex:counter} (counter).}
	\label{fig:counter-verilog}
\end{figure}
We apply $\tool$ to traces generated from the following hardware designs.
Note that, since $\tool$ observes traces and treats the system that generates the traces as a black box, the performance of $\tool$ does not depend on
the size of the circuit.

\begin{example}[\textsc{xor}]
  As a first example, consider the \textsc{xor} function $\vec{o} = \vec{i} \oplus \vec{i}'$.
  In the corresponding circuit, every $j$-th output bit $o_j$ is only influenced by the $j$-the input bits $i_j$ and $i'_j$.
\end{example}

\begin{example}[\textsc{mux}]
This example circuit is depicted in Figure~\ref{fig:example-circuit}.
There is a black box combinatorial circuit, guarded by a multiplexer that selects between the two input vectors $\vec i$ and $\vec i'$ and an inverse multiplexer that forwards the output of the black box either towards $\vec o$ or $\vec o'$.
Despite there being a syntactic dependency between $\vec o$ and $\vec i'$, there is no semantic dependency, i.e., the output $\vec o$ does solely depend on $\vec i$ and the selector signal.

When using the same example, but with a sequential circuit as black box, there may be information flow from the input vector $\vec i'$ to the output vector $\vec o$ because the state of the latches may depend on it.
We construct such a circuit that leaks information about $\vec{i}'$ via its internal state.

\end{example}

\begin{example}[counter]
	\label{ex:counter}
	Our last example is a binary counter with two input control bits $\mathit{incr}$ and $\mathit{decr}$ that increments and decrements the counter.
	The corresponding Verilog design is shown in Figure~\ref{fig:counter-verilog}.
	The counter has a single output, namely a signal that is set to one when the counter value overflows.
	Both inputs influence the output, but timing of the overflow depends on the number of counter bits.
\end{example}

\begin{table}[t]
  \caption{Results of $\tool$ on traces generated from circuit instances. Every instance was run 10 times with different seeds and the average is reported.}
  \label{tbl:rvhyper-results}
  \centering
  \begin{tabular}{llllllll}
  \hline \noalign{\smallskip}
  instance & property & satisfied & \#\,traces & length & time & \#\,instances \\ \noalign{\smallskip}\hline\noalign{\smallskip}
      \textsc{xor} & $i_0 \ninfluences o_0$ & no & 18 & 5 & 12ms & 222  \\
      \textsc{xor} & $i_1 \ninfluences o_0$ & yes & 1000 & 5 & 16913ms & 499500  \\
      counter & $\textit{incr} \ninfluences \textit{overflow}$ & no & 1636 & 20 & 28677ms & 1659446 \\
      counter & $\textit{decr} \ninfluences \textit{overflow}$ & no & 1142 & 20 & 15574ms & 887902 \\
  \textsc{mux} & $\vec i' \ninfluences \vec o$ & yes & 1000 & 5 & 14885ms & 499500 \\
      \textsc{mux2} & $\vec i' \ninfluences \vec o$ & no & 82 & 5 & 140ms & 3704 \\ \hline
  \end{tabular}
\end{table}

\subsubsection{Results.}
The results of multiple random simulations are given in Table~\ref{tbl:rvhyper-results}.
Despite the high complexity of the monitoring problem, $\tool$ is able to scale up to thousands of input traces with millions of monitor instantiations (cf.~Algorithm~\ref{alg:offline}).
$\tool$'s optimizations, i.e., keeping only a minimal set of traces and reducing the number of instances by the specification analysis, are a key factor to those results.
For the two instances where the property is satisfied (\textsc{xor} and \textsc{mux}), $\tool$ has not found a violation for any of the runs.
For instances where the property is violated, $\tool$ is able to find counter examples.
While counter examples can be found quickly for \textsc{xor} and \textsc{mux}2, the counter instances need more traces since the chance of finding a violating pair of traces is lower.

\section{Conclusion}

RVHyper monitors a running system for violations of a HyperLTL specification. The functionality
of RVHyper thus complements model checking tools for HyperLTL, like MCHyper~\cite{conf/cav/FinkbeinerRS15}, and tools for satisifability checking, like EAHyper~\cite{conf/cav/FinkbeinerHS17}. RVHyper is in particular useful during the development of a $\hyperltl$ specification, where it can be used to check the HyperLTL formula on sample traces without the need for a complete model.
Based on the feedback of the tool, the user can refine the HyperLTL formula until it captures the intended policy.


\bibliographystyle{splncs03}
\bibliography{main}



\end{document}